\documentclass[aps,prl,preprint,showpacs,superscriptaddress]{revtex4}
\usepackage{amsmath}
\usepackage{amsfonts}
\usepackage{amssymb}%
\usepackage{graphicx,epsfig}
\begin{document}
\title{Tuning the metamagnetism of an antiferromagnetic metal}

\author{J. B. Staunton}
\affiliation{Department of Physics, University of Warwick, Coventry CV4 7AL, U.K.}
\author{M. dos Santos Dias}
\affiliation{Department of Physics, University of Warwick, Coventry CV4 7AL, U.K.}
\affiliation {Peter Gr\"unberg Institut and Institute for Advanced Simulation, Forschungszentrum J\"ulich and JARA,
D-52425 J\"ulich, Germany}
\author{J. Peace}
\affiliation{Department of Physics, University of Warwick, Coventry CV4 7AL, U.K.}
\author{Z. Gercsi}
\affiliation{Department of Physics, Blackett Laboratory, Imperial College London, SW7 2AZ, U.K.}
\author{K. G. Sandeman}
\affiliation{Department of Physics, Blackett Laboratory, Imperial College London, SW7 2AZ, U.K.}

\date{\today}

\begin{abstract}
We describe a `disordered local moment' (DLM) first-principles electronic structure theory which demonstrates that tricritical metamagnetism can arise in an antiferromagnetic metal due to the dependence of local moment interactions on the magnetisation state.    Itinerant electrons can therefore play a defining role in metamagnetism in the absence of large magnetic anisotropy.  Our model is used to accurately predict the temperature dependence of the metamagnetic critical fields in CoMnSi-based alloys, explaining the sensitivity of metamagnetism to Mn-Mn separations and compositional variations found previously.  We thus provide a  finite-temperature framework for modelling and predicting new metamagnets of interest in applications such as magnetic cooling.  

\end{abstract}
\maketitle
The application of a magnetic field to an antiferromagnet can cause abrupt changes to its magnetic state~\cite{stryjewski_1977a}.  While such metamagnetic transitions have been known and studied for a long time in materials such as FeCl$_2$~\cite{jacobs_1967a} and MnF$_2$~\cite{jacobs_1964a}, it is their association with technologies such as magnetic cooling~\cite{sandeman_2012a} that has driven recent efforts to control metamagnetism in the room temperature range, and at accessible magnetic fields. 

In the particular case of magnetic cooling, a large, or even `giant' magnetocaloric effect can arise at a first order 
metamagnetic phase transition such as is found in FeRh~\cite{annaorazov_1992a}.  However hysteresis can be present at these 
transitions, resulting in significant inefficiency during magnetic cycling.  Therefore, antiferromagnetic (AFM) metamagnets 
that sit on the border between first order and second order behavior are of great interest. They emulate the low 
hysteresis, high entropy change properties of two of the most advanced magnetic refrigerants, La-Fe-Si~\cite{Dung} and 
MnFe(P,As)~\cite{Liu}, both of which are instead ferromagnetic (FM) metamagnets that have field-induced metamagnetic 
critical points.  It is worth noting that tricritical points have also been examined in the context of other technologies, such as displays based on 
liquid crystals~\cite{C+L}.

Metamagnetic transitions have previously been investigated principally in terms of either localised or itinerant electron spin effects. In antiferromagnetic insulators~\cite{C+L}, magnetic field-driven phase transitions can be understood qualitatively using a localised (classical) spin 
Hamiltonian with simple pairwise isotropic exchange interactions, a source of magnetic anisotropy and a Zeeman term for the 
influence of an external magnetic field.  For example, Nagamiya~\cite{Nagamiya} showed that, if the localised spins of a 
helical antiferromagnet are pinned by anisotropy and crystal field effects to spiral around a particular direction, the 
effect of a magnetic field is to bring about a first order transition to a fan structure where the moments now oscillate 
about the field direction.  At higher fields, the fan angle smoothly reduces to zero to establish a high magnetisation 
phase at a second order transition. However, if anisotropy effects are negligible, there is now no favored axis for the 
helical order and, as soon as a magnetic field is applied, the helix plane orients perpendicular to the field. As the field is increased the moments smoothly cant into a conical spiral towards the field's direction with a second order transition to a high magnetisation phase.  Thus in this localised picture, first order metamagnetism relies on a source of anisotropy.

Metamagnetism is accounted for differently in an itinerant electron system.  Seminal work by Wohlfarth and Rhodes~\cite{W+R}, Moriya and Usami~\cite{M+U}, developed and extended by many others, e.g.~\cite{Shimuzu,Yamada}, derived the coefficients of a Landau-Ginzburg expansion of the free energy of an
antiferromagnet exposed to a uniform magnetic field in terms of the antiferromagnetic order parameter, $\Delta m_{\vec{q}}$, with wave-vector modulation $\vec{q}$.  The coefficients of the expansion are determined from a consideration of both Stoner particle-hole excitations~\cite{W+R} and spin fluctuations~\cite{M+U,Yamada} generated from the collective behavior of the interacting electrons~\cite{SFT,M+U,Kubler-book}.  Field-induced tricriticality occurs when the quadratic and quartic terms in  $\Delta m_{\vec{q}}$ both equal zero. More recently, from an analysis of a generic mean-field Hamiltonian describing a helical antiferromagnetic state in an applied field Vareogiannis~\cite{Vare} has shown how an itinerant electron system can undergo a first-order 'spin-flip' transition. For a weak magnetic field the AFM polarization is parallel to the applied field and flips perpendicular only when the field exceeds a critical value~\cite{Irisawa}. 

In this Letter we describe an ab-initio spin density functional theory (SDFT)-based `local moment' theory for antiferromagnetic metals that blends these two complementary accounts of metamagnetism to suit materials where slowly varying 'local moments' can be identified from the complexity of the electronic behavior.  A principal result is that, in the limit where magnetic anisotropy effects are small or neglected entirely, the local moments' antiferromagnetic order can still undergo a first order transition to a fan or ferromagnetic state owing to the feedback between the local moment and itinerant aspects of the electronic structure.  We test our theory against a detailed experimental case study of CoMnSi-based tricritical metamagnets and show how it provides quantitative materials-specific guidance for tuning metamagnetic and associated technological properties.  

We start from a generalisation of SDFT~\cite{blg} that describes the `local moment' picture of metallic magnets at 
finite temperature \cite{jh,hh,JBS+BLG,RDLM-FePt}. Its basic premise is a timescale separation between fast and slow 
electronic degrees of freedom so that `local moments' are set up with slowly varying orientations, $\{\hat{e}_i\}$. The 
existence and behavior of these `disordered local moments' (DLM)  is established by the fast electronic motions and 
likewise their presence affects these motions. This mutual feedback can then lead to a dependence of the local moments' 
interactions on the type and extent of the long range magnetic order through the associated itinerant electronic structure.  The DLM picture  of the paramagnetic state can be mapped to an Ising picture with, on average, one half of the moments oriented one way and the rest antiparallel. However once the symmetry is broken so that there is a finite order parameter $ \{ \vec{m}_i \}$ profile, e.g in a FM or AM state and/or when an external magnetic field is applied, this simplicity is lost. Ensemble averages over the full range of non-collinear local moment orientational configurations, $\{\hat{e}_i\}$, are needed to determine the system's magnetic properties realistically ~\cite{RDLM-FePt}.  We now develop this idea and extend the DLM theory to enable the effects of an external magnetic field to be modelled along with temperature and field dependence of the magnetic state.

We consider a magnetic material in an external magnetic field $\vec{B}$ at a temperature $T$. The
probability that the system's local moments are configured according to  
$\{ \hat{e}_i \}$ 
is $ 
P_i ( \{ \hat{e}_i \} ) = \exp [ - \beta \Omega 
( \{ \hat{e}_i \},\vec{B} ) ]/Z$
where the partition function  
$Z =$ $ \prod_j \int d \hat{e}_j \exp [ - \beta \Omega
 ( \{ \hat{e}_i \} ,\vec{B}) ] $ and $ \beta = 
(k_B T ) ^{-1}$. An ab-initio `generalised' electronic grand potential
$ \Omega ( \{ \hat{e}_i \} ,\vec{B}) $  is in principle available from SDFT~\cite{blg} where the spin density is constrained to be orientated according to the local moment configuration $\{ \hat{e}_i \}$. The thermodynamic free energy, which accounts for both the entropy associated with the orientational fluctuations of the local moments as well as that for the remaining electronic degrees of freedom, is given by $F= -k_{B} T \log Z$. $\Omega (\{\hat{e}_{i}\},\vec{B})$ thus plays the role of a local moment Hamiltonian but its genesis can give it a complicated form. Nonetheless by expanding about a suitable reference `spin' Hamiltonian
$\Omega_{0}\{ \hat{e}_{i}\} = \sum_{i} \vec{h}_i \cdot \hat{e}_i$
and, using the Feynman Inequality~\cite{Feynman}, we can find a mean field theoretical estimate of the free energy~\cite{blg}
\begin{equation}
\label{Free}
\begin{split}
F (\{ \vec{m}_i \},\vec{B},T) = \langle \Omega(\{\hat{e}_{i}\},\vec{B}) \rangle_{\{\vec{m}_i\}} \\
 + k_B T \sum_i
\int P_i (\hat{e}_i) \ln P_i (\hat{e}_i) d\hat{e}_i  -\vec{B} \cdot \sum_i \mu_i \vec{m}_i
\end{split}
\end{equation}
where the probability of a moment pointing along $\hat{e}_i$ on a site $i$ is
$ P_i (\hat{e}_i) = \exp[-\beta \vec{h}_i
 \cdot \hat{e}_i]/\int \exp[-\beta \vec{h}\cdot \hat{e}_i]  d \hat{e}_i$, 
so that the set of local order parameters, each of which can take values between $0$ and $1$, $\{\vec{m}_i\} =\{\int \hat{e}_i P_i (\hat{e}_i)d \hat{e}_i\} 
=\{\langle \hat{e}_i \rangle \}$. Our DLM self-consistent mean-field theory of the statistical mechanics of the
local moments can be seen as the natural counterpart of the DFT self-consistent description of the interacting electrons. The first term of Eq.~\ref{Free} is the internal energy of the local moments, i.e.
the average of the electronic grand potential over local moment configurations consistent with the order parameter profile 
$\{\vec{m}_i\}$, the second is $(-T)$ multiplied by the local moments' entropy and the last their interaction with a field, $\vec{B}$. 
The sizes of the local moments, $\{ \mu_i\}$, are determined self-consistently~\cite{blg} via the generalised SDFT.

The Weiss field at a site $l$ is given by
\begin{equation}
\label{weiss}
\vec{h}_l =- \frac{ \partial \langle \Omega  (
\{ \hat{e}_i \} ,\vec{B})\rangle_{\{\vec{m}_i\}}}{\partial \vec{m}_l}.
\end{equation}
To capture 
the itinerant electronic component of the problem coming from the overall spin-polarisation of the electronic structure,
we approximate the $\vec{h}_l$ as the first two terms of
an expansion about a uniform distribution $\vec{m}= \frac{1}{N} \sum_i \vec{m}_i$, i.e.
\begin{eqnarray}
\label{expand}
\vec{h}_l & \approx &- \left. \frac{ \partial \langle \Omega \rangle  }{\partial \vec{m}_l}\right |_{\vec{m}} -
\sum_{j} \left. \frac{\partial^2 \langle \Omega \rangle}{\partial \vec{m}_{l} \partial \vec{m}_{j}} \right |_{\vec{m}}
\cdot 
(\vec{m}_{j} - \vec{m}) \\ \nonumber
&=& \vec{h}(\vec{m}) +\sum_{j} \tilde{S}^{(2)}_{l,j}(\vec{m}) \cdot 
(\vec{m}_{j} - \vec{m})
\end{eqnarray}
where $\tilde{S}^{(2)}_{i,j}(\vec{m})$ is the direct correlation function tensor and describes effective interactions between 
the local moments that depend on the magnitude of long range magnetic order, $\vec{m}$. 

A solution of Eqs.\ref{Free} to \ref{expand} at a fixed $T$ and applied field $\vec{B}$ minimises the free energy, Eq.~\ref{Free}.  Several solutions, $\{ \vec{m}^{(1)}_l \}$,$\{ \vec{m}^{(2)}_l \}$,$\cdots$ 
may be found and the one which produces the lowest free energy, Eq.~\ref{Free}, describes the equilibrium state of the 
system, $\{ \vec{m}_l \}_{Equil.}$. Hence metamagnetic transitions can be tracked as functions of $T$ and $\vec{B}$ -
for a given $T$ the solutions for increasing values of $B$ can show a transition from, say, an antiferromagnetic to 
ferromagnetic state at a critical field, $B_c$ (e.g. see Fig.~1(b)). The material's spin-polarised electronic structure 
also depends on $B$ and $T$ and the state of magnetic order. 

We now follow through the general framework laid out in Eqs.\ref{Free} to \ref{expand} for a putative helical 
antiferromagnetic metal. In order to elucidate the main aspects, we shall assume temporarily that magnetic anisotropy 
effects are small and can be neglected. In the absence of an applied magnetic field, the solution $\{ \vec{m}_l \}$ which 
produces the  lowest free energy has the form $\vec{m}_i = \Delta m_{\vec{q}} ( \cos (\vec{q} \cdot \vec{R}_i) \hat{x} + 
\sin (\vec{q} \cdot \vec{R}_i) \hat{y})$. The helical axis, $\hat{z}$, has no preferred direction in the crystal lattice. The order parameter $\Delta m_{\vec{q}}$ increases from $0$ to $1$ as $T$ drops from $T_N$ to $0$~K. On application of a magnetic field, $\vec{B}$, defining, say, the x-axis of our coordinate frame, we can find numerical solutions of the mean field equations \ref{Free} to \ref{expand} of (i) distorted helical form:
$\vec{m}_i =(m+\Delta m_{\vec{q}})  \cos (\vec{q} \cdot \vec{R}_i) \hat{x} + \Delta m_{\vec{q}}
\sin (\vec{q} \cdot \vec{R}_i) \hat{y}$, (ii) a conical helix: $\vec{m}_i =m \hat{x} + \Delta m_{\vec{q}}  
(\cos (\vec{q} \cdot \vec{R}_i) \hat{y} + \sin (\vec{q} \cdot \vec{R}_i) \hat{z})$, (iii) a fan state:
$\vec{m}_i = m \hat{x} + \Delta m_{\vec{q}}  \cos (\vec{q} \cdot \vec{R}_i) \hat{y}$ and (iv) a ferromagnetic state with
$\Delta m_{\vec{q}}=0$. The free energies are $F_{dh}(m,\Delta m_{\vec{q}}; B,T)$, $F_{ch}(m,\Delta m_{\vec{q}}; B,T)$, 
 $F_{fan}(m,\Delta m_{\vec{q}}; B,T)$ and $F_{FM}(m ; B,T)$ respectively and we search for the lowest one at a given $B$ and 
$T$.  Consequently the relative difference between $F_{dh}$ and any of the others determines whether there is a first order metamagnetic 
transition or not and if so the temperature dependence of the critical field, $B_c$. 

The local moment interactions $\tilde{S}^{(2)}_{i,j}(\vec{m})$ of Eq.~\ref{expand} set up the Weiss field and, along with 
the uniform component, $\vec{h}(\vec{m})$, are the key elements of our theory.  We determine these quantities ab-initio using 
relativistic, spin-polarised, multiple scattering (Korringa-Kohn-Rostoker, KKR) theory and the coherent potential 
approximation (CPA) to handle the electrons moving in fields set by the orientationally disordered local 
moments~\cite{RDLM-FePt,RDLM-big,Manuel}. Here, 
for the first time, we account for the variation of $\tilde{S}^{(2)}_{i,j}(\vec{m})$ with increasing extent of 
long-range magnetic order $\vec{m}$ driven by spin-polarisation of the itinerant electrons which mediate the interaction between the local 
moments~\cite{Sandratskii,Polesya}.  Since spin-orbit coupling effects are also included~\cite{Manuel} we can assess 
whether the neglect of magnetic anisotropic effects is reasonable. Details on how  $\tilde{S}^{(2)}_{i,j}(\vec{m})$ is calculated can be found in references 
\cite{blg,Manuel,Nature,TM-oxides}.

Our previous experimental investigations have revealed a class of magnetic materials based on the orthorhombic CoMnSi 
metallic antiferromagnet to be an ideal testing ground for the theory~\cite{IC1,IC2,IC-2ndpaper}. Those studies considered 
both the composition-dependent metamagnetism and pronounced magneto-elasticity in CoMnSi by extensive magnetic and 
structural measurements, including a characterisation of the anomalous temperature variation of structural parameters in 
zero magnetic field.  CoMnSi orders into an non-collinear, helical anti-ferromagnetic state in zero magnetic field 
at $T_N \approx $380~K. In an applied field this transition becomes a metamagnetic one to a high magnetisation/ferromagnetic
 state at a temperature $T_t$. As the applied field is increased $T_t$ decreases before going through a tricritical point 
at around 2 Tesla where an enhanced magnetocaloric effect is observed along with marked changes to the interatomic spacings.
All three subsystems, Co, Mn and Si can be doped and substituted which also affects the magnetic 
properties~\cite{IC-2ndpaper}. This material is typical of many useful magnetic metals in that its magnetism possesses both localised and itinerant electron spin attributes~\cite{Irisawa}. Localised magnetic moments can be identified with the Mn sites whereas the magnetism associated with the Co sites is reflected in the long range spin polarisation of the electronic structure. Our measurements have also found that the magnetism of CoMnSi is dominated by the behavior of the Mn moments. Their interactions, however, are delicately poised and depend on the spacing between them, the compositional environment in which they sit and the overall long range magnetisation. 

We previously modelled a commensurate approximation to the helical AFM state~\cite{IC2} found in experiment and other calculations have examined an assumed 
FM state~\cite{LTP}.  This Letter makes the first ab-initio exploration of the non-collinear helical magnetism of CoMnSi and its magnetic field and 
temperature variation. We start with self-consistent KKR-CPA calculations~\cite{KKRCPA,scf-kkr-cpa,KKR-review} 
for the paramagnetic ($m=0$) DLM state of CoMnSi.  Local moments of magnitude $\mu \approx$ 3.0$\mu_B$ are established
 on the Mn sites which are very close to the magnetisation per Mn site we find in calculations of CoMnSi in a FM 
state ($m=1$) and we can assume that the sizes of the Mn local moments are rather insensitive to the orientations of 
moments surrounding them. We thus use the potentials within the 'frozen potential' approximation~\cite{frozen,RDLM-big} to study CoMnSi for probability distributions
$\{P_i(\hat{e}_i)\}$ with a set of values of the magnetic order parameter $\vec{m}_i$, each ranging between $0$ and $1$. We use an 
angular momentum cut-off of $l=$ 3, a muffin-tin approximation for the potentials and local spin density approximation for 
exchange and correlation.  For our DLM calculations for a finite long range order parameter, $m \neq 0$, we consider a 
probability distribution on a fine grid of local moment orientations
~\cite{RDLM-FePt,RDLM-big}. No local moment forms on the cobalt sites in the DLM paramagnetic state whereas for finite 
$m$, a small magnetisation associated with each Co site is induced by the lining up of the Mn moments and the consequent 
overall spin polarisation of the electronic structure. For the FM state, $m =1$, this is $\approx 0.6\mu_B$ per 
Co site~\cite{IC2,LTP}.

The onset of magnetic order in CoMnSi and the temperature at which this transition takes place is obtained from an 
examination of $\tilde{S}^{(2)}_{i,j}(\vec{m})$ for the paramagnetic state, $m=0$. By calculating the lattice Fourier 
transform of this interaction, $\tilde{S}^{(2)}(\vec{q},0)$ and finding the wave-vector, $\vec{q}_{max}$, where the largest
 eigenvalue, $s^{(2)}(\vec{q},0)$, maximises, we find the magnetically ordered state that the system forms below our 
mean-field theory estimate of $T_{N}=s^{(2)} (\vec{q}_{max},0)/3k_B$. If $q_{max}=0$ a FM state is indicated whereas 
$q_{max}\ne 0$ points to an AF state. Fig.1(a) shows $s^{(2)}(\vec{q},0)$ using structural data obtained by neutron diffraction~\cite{IC1}. These data show the Mn-Mn
distances, $d_1$, to vary by more than 2$\%$ over the temperature range 100-400K. Our calculations show that CoMnSi should 
order into an incommensurate helical AFM state along the c-axis, set by the orthorhombic crystal structure, at $T_N \approx 400$K in good agreement with experiment~\cite{IC1,IC2}.  They confirm the expectation that spin-orbit coupling effects are small in CoMnSi, with only a very weak anisotropy favoring the Mn moments to lie in the $a$-$b$ plane.  This agrees with the very small level of magnetic anisotropy seen in measurement~\cite{morrison_2008a}.

CoMnSi is evidently very close to a ferromagnetic instability as shown by the value of $s^{(2)} (\vec{q}=0,\vec{m}=0)$ 
being close to that of $s^{(2)} (\vec{q}_{max},\vec{m}=0)$, a convenient signature for a potentially useful metamagnet. Its
marked magnetoelastic properties are confirmed by the tendency to order ferromagnetically taking over from the 
incommensurate ordering propensity as $d_1$ is increased, which experiment also shows to happen with increase in 
temperature.  When the bath of electrons in which the Mn local moments sit develops a long range spin polarisation as a
 magnetic field is  applied and $m$ increases, our calculations show that the magnetic interactions, $\tilde{S}^{(2)}(\vec{q},\vec{m})$, 
weaken significantly only strengthening again in the spin-wave limit of $m\approx 1$. 
The inset to Fig.~1(a) illustrates this.  The magnetic anisotropy effects remain very small,  ( e.g we find a uniaxial anisotropy $K < $ 0.01mRy. per unit cell for the completely ferromagnetically ordered state with a small orbital moment of $<$ 0.03$\mu_B$ on the Mn sites).  

\begin{figure}[h]
\begin{center}
\includegraphics[width=40mm,height=35mm]{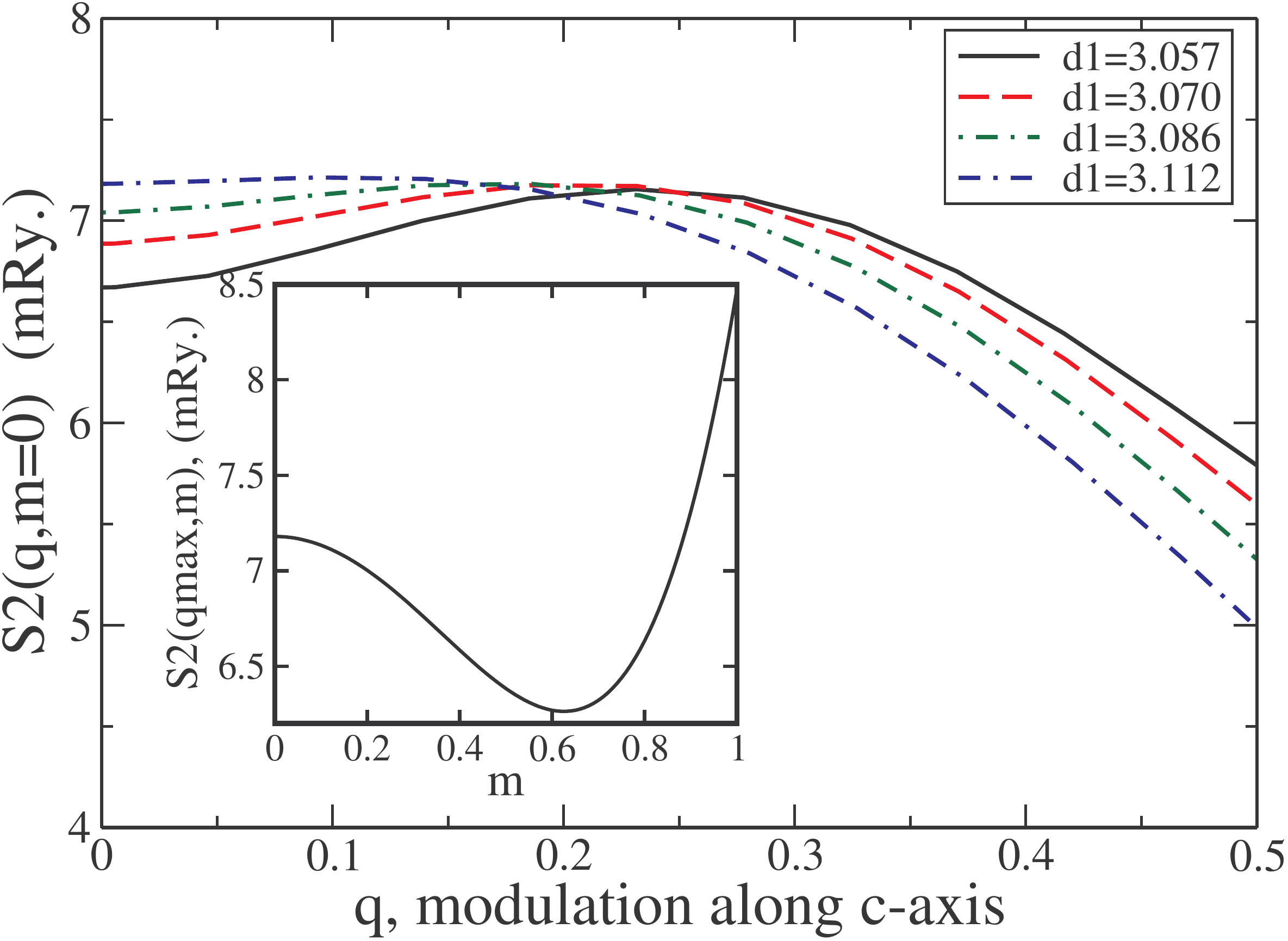}
\includegraphics[width=40mm,height=35mm]{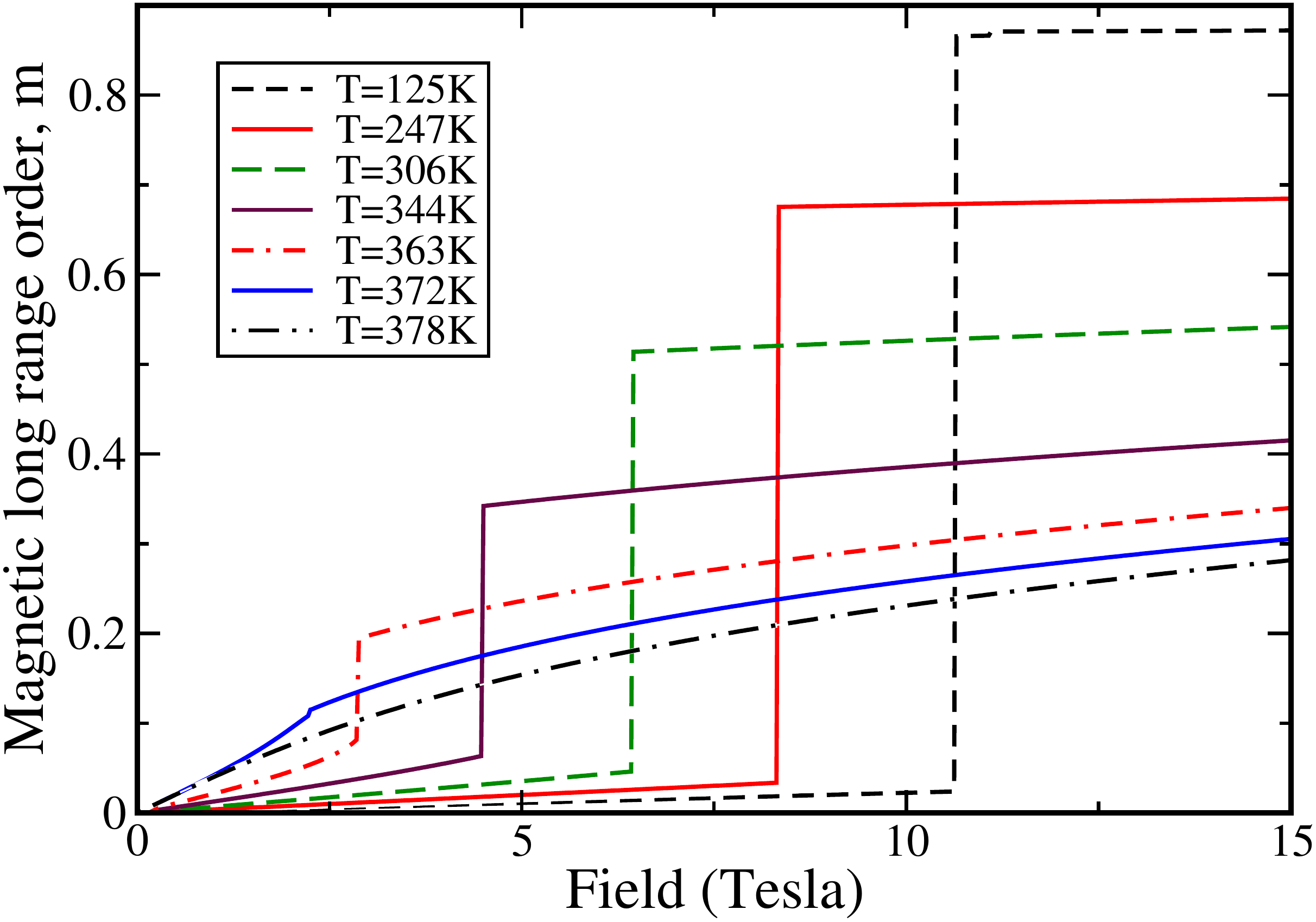}
\caption{(a) The Mn local moment interactions, $s^{(2)} (\vec{q}, \vec{m}=0)$, for CoMnSi with
structures measured in neutron diffraction experiments at four temperatures labelled by the Mn-Mn spacing, $d_1$, in \AA. 
The four Mn moments within a unit cell tend to align with each other. The inset shows 
the variation of $s^{(2)} (\vec{q}_{max}, \vec{m})$ with FM order parameter $m$ for the case where $d_1=$
 3.07~\AA. (b) The variation of  $m$ versus applied field for several 
temperatures for a fixed lattice structure, in which $d_1=$ 3.07~\AA. A tricritical point is indicated at 372K, 2 Tesla.}
\label{Fig1}
\end{center}
\end{figure}

The ferromagnetically-induced weakening of the magnetic interactions acts to promote the stability of the 
distorted helical magnetic state over other states when the magnetic field is applied~\cite{LTP}. There can therefore be a first order transition to a 
fan, conical helix or ferromagnetic state at critical field $B_c$ despite the absence of magnetic anisotropy.  Fig.~1(b) 
illustrates this effect, and shows our solutions of the mean field theory equations \ref{Free} to \ref{expand} where we 
have used our calculated ab-initio uniform Weiss field $\vec{h}(\vec{m})$~\cite{RDLM-FePt} and 
$\tilde{S}^{(2)}(\vec{q}_{max},\vec{m})$ values. We point out that there are no adjustable parameters for these curves which show a transition
from a distorted helical AF state (small $m$, $B$) to a high magnetisation state above a critical field $B_c$. The transition
is first order up to a tricritical temperature of 372K and second order thereafter.

Fig.~2 shows effect on the temperature dependence of the critical field, $B_c$ of either changing Mn nearest-neighbor 
separations, $d_1$, (Fig.~2(a)) in CoMnSi or varying composition slightly away from CoMnSi (Fig.~2(b)). If the itinerant 
electron effect which gives   $\tilde{S}^{(2)} (\vec{q},\vec{m})$ its $\vec{m}$-dependence is neglected and only the 
effects of a weak on-site magnetocrystalline anisotropy are included, we find the values of $B_c$ to be 2 orders of 
magnitude smaller. $B_c$ increases sharply as $d_1$ decreases. This is in line with measured pressure dependence of $B_c$ of
CoMnSi~\cite{JMMM}. $B_c$ is also very sensitive to compositional doping -
the AFM tendency strengthens as electrons are removed so that $B_c$ increases, as shown by an application to CoMn$_{95}$Cr$_{5}$Si, 
whereas it weakens when electrons are added, as in the case of Co$_{95}$Ni$_{5}$MnSi. The CPA is used to describe the compositional doping on the Co and Mn sublattice as well as the local moment disorder on the Mn sites. 

\begin{figure}[h]
\begin{center}
\includegraphics[width=40mm,height=35mm]{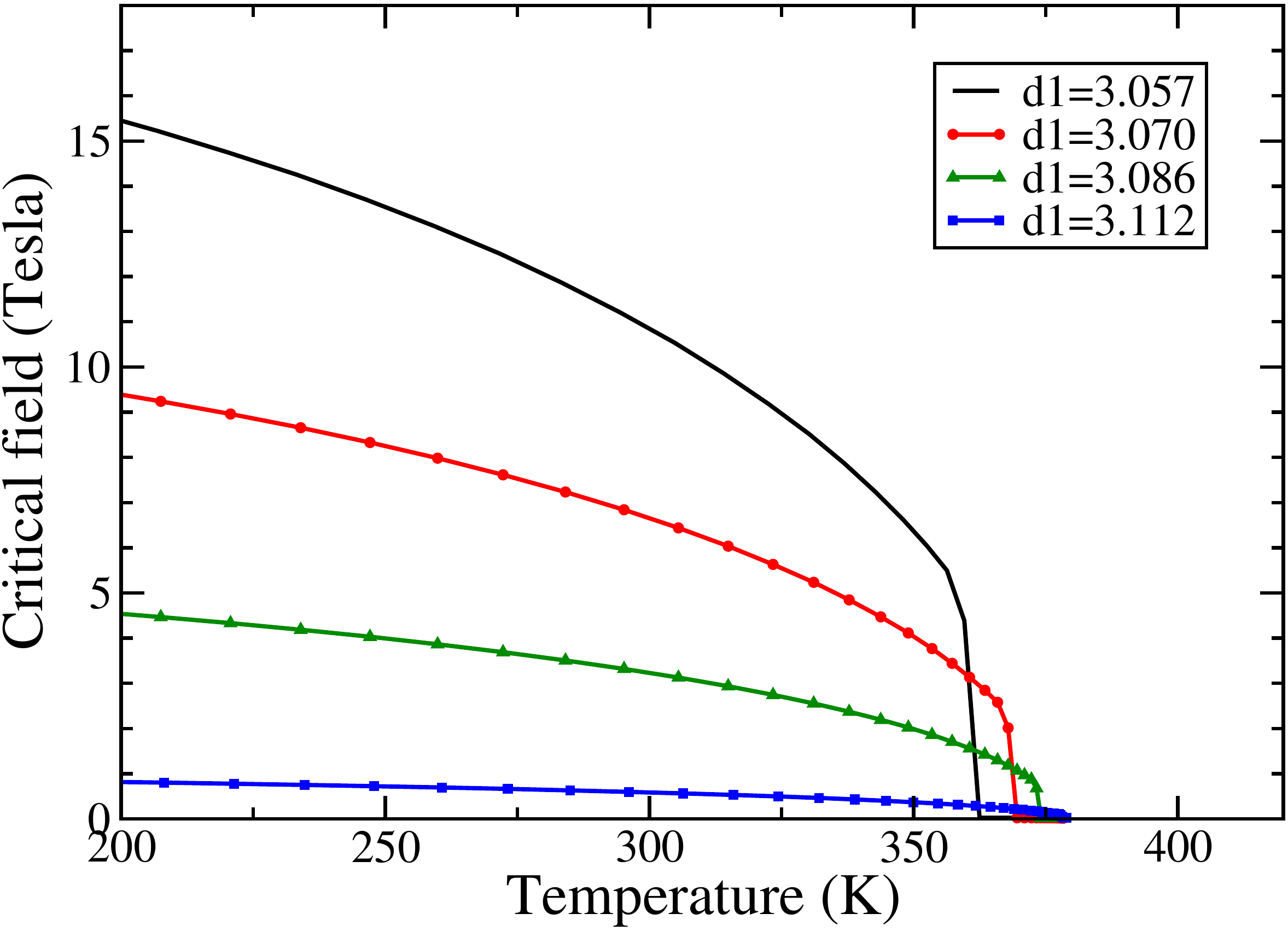}
\includegraphics[width=40mm,height=35mm]{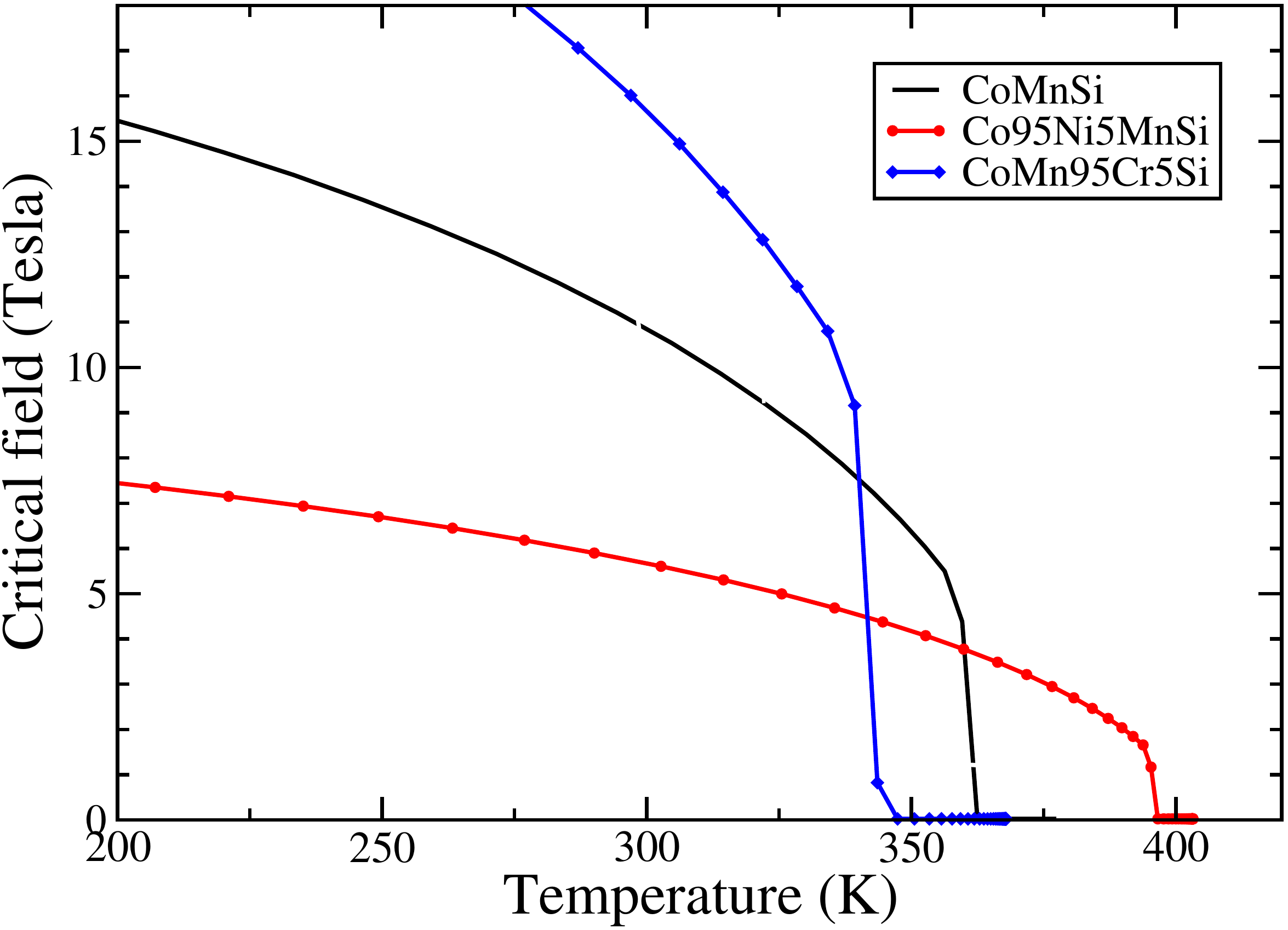}
\end{center}
\caption{ (a) The critical field $B_c$ for transition between a helical AF and ferromagnetic state versus $T$ for 
CoMnSi for the four structures referred to in Fig.1. and (b)  for Co$_{95}$Ni$_{5}$MnSi, and  CoMn$_{95}$Cr$_{5}$Si,  with $d_1=$ 3.06~\AA.}
\label{Fig2}
\end{figure} 

To test the theory presented here we have carried out further analysis of our experimental data to extract the temperature 
dependence of the critical magnetic fields of CoMnSi and its compositionally doped counterparts, Co$_{95}$Ni$_{5}$MnSi 
and CoMn$_{98}$Cr$_{2}$Si. These results are shown in Fig.~3(a).
\begin{figure}[h]
\begin{center}
\includegraphics[width=70mm,height=60mm,angle=-90]{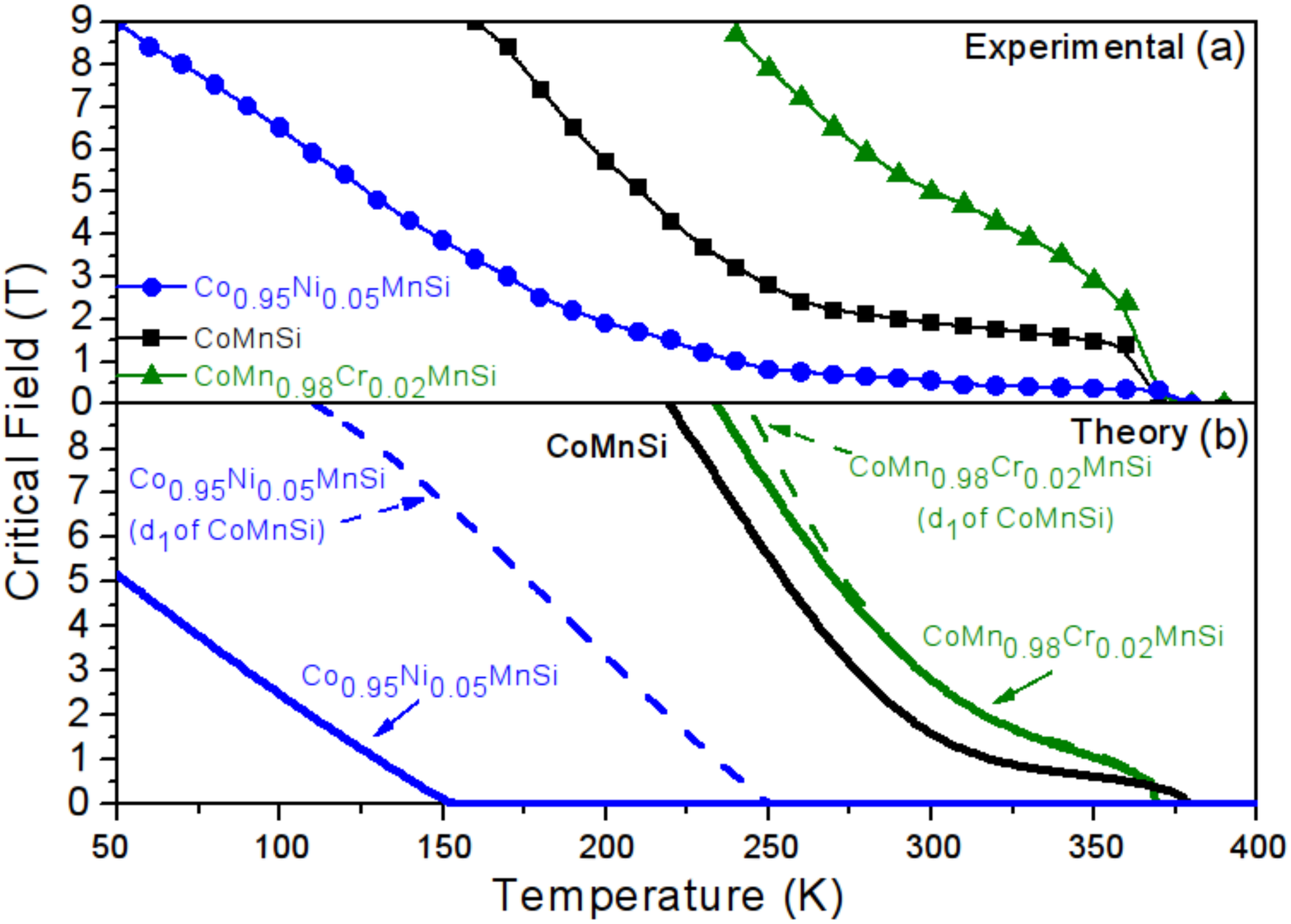}
\end{center}
\caption{ (a) The critical field $B_c$ versus $T$ for CoMnSi, Co$_{95}$Ni$_{5}$MnSi and 
CoMn$_{98}$Cr$_{2}$Si measured in experiment.  (b) $B_c(T)$ for the same three 
systems from theory, using the experimentally determined lattice structures~\cite{IC-2ndpaper} in zero field. The Ni- and 
Cr-doped systems are also shown using the CoMnSi structural data (dashed lines).}
\label{Fig3}
\end{figure}
The lower panel, Fig.~3(b) shows our theoretical estimates of the critical fields for these same three systems, where we 
have used the zero field experimental structural data. Overall the magnitudes and trends are given very well by the theory. 
The variation in temperature of the relative spacings between the Mn atoms is responsible for concave appearance of 
$B_c(T)$ in CoMnSi. The Ni-doped structure is shown to be extremely sensitive to structure, as observed 
experimentally~\cite{IC-2ndpaper}.  When $B_c >$ 2 Tesla, $\frac{dB_c}{dT}$ appears to be overestimated by the theory. This discrepancy could be partly mitigated by allowing for the magnetic-field induced increase in $d_1$ which can inferred from our calculations of $\vec{h}(\vec{m})$ and $\tilde{S}^{(2)}(\vec{q}_{max},\vec{m})$ which show that ferromagnetic tendency strengthens with increase in $d_1$.

Detailed theoretical models are needed to use in concert with experimental studies if the behavior of highly sensitive 
magnetic materials is to be analysed and tuned. We have discussed and demonstrated
one such model here and showed how the temperature and field dependence of the metamagnetic transition of an 
antiferromagnetic metal can be affected by varying composition and atomic spacing. Useful magnetic metals typically have 
large magnetic moments which are established locally. The Disordered Local Moment model works well for 
these~\cite{TM-oxides, RDLM-FePt, Nature} and we have set out a theory for metamagnetic transitions based on this picture. 
An important aspect for magnetic material design is to distinguish first and second order magnetic transitions and to get 
some control of the location of precious tricritical points. Our CoMnSi case study reveals how reducing the Mn-Mn 
separation, $d_1$,  increases $B_{c}$. Small degrees of compositional doping which reduce the number of electrons have a similar effect. One notable feature in our modelling is the dependence of the interaction between the local moments upon the extent of long range magnetic order. This comes from the change in the behavior of the itinerant electrons that mediate these interactions and produces a mechanism for first-order `spin-flip' transitions even for cases where there is little magnetic anisotropy.  This aspect has an important role in the analysis and design of adaptive, magnetic metals.  

We acknowledge financial support from the EPSRC (UK) (J.B.S, J.P), a FCT Portugal PhD grant SFRH/ BD/35738/2007 (M.d-S.D), 
the Royal Society (K.G.S) and EPSRC grant EP/G060940/1 (Z.G).  K.G.S. thanks M. Avdeev and J. Bechh\"ofer for useful discussions.


\begin{thebibliography}{99}
\bibitem{stryjewski_1977a} E. Stryjewski and N. Giordano, Adv. Phys. {\bf 26} 487 (1977).
\bibitem{jacobs_1967a} I. S. Jacobs and P. E. Lawrence, Phys. Rev. {\bf 164} 866 (1967).
\bibitem{jacobs_1964a} I. S. Jacobs and P. E. Lawrence, J. Appl. Phys. {\bf 35} 996 (1964).
\bibitem{sandeman_2012a} K.G. Sandeman, Scripta Materialia {\it in press} (2012) doi: 10.1016/j.scriptamat.2012.02.045
\bibitem{annaorazov_1992a} M.P. Annaorazov et al., Cryogenics {\bf 32} 867 (1992).
\bibitem{Dung}N.H. Dung et al., Adv. Energy Mater. 1, 1215 (2011).
\bibitem{Liu} J. Liu et al., Scripta Mater. (in press, 2012) doi: 10.1016/j.scriptamat.2012.02.039
\bibitem{C+L} P. M. Chaikin and T. C. Lubensky, {\it Principles of condensed matter physics}, C.U.P. (2000).
\bibitem{Nagamiya} T. Nagamiya et al., Prog. Theor. Phys. {\bf 27}, 1253, (1962).
\bibitem{W+R} E. P. Wohlfarth and P. Rhodes, Phil. Mag. {\bf 7}, 1817, (1962).
\bibitem{M+U} T. Moriya and K. Usami, Sol. Stat. Comm. {\bf 23}, 935, (1977).
\bibitem{Shimuzu} M. Shimuzu, J. Physique {\bf 43}, 155, (1982).
\bibitem{Yamada} H. Yamada and T. Goto, Phys. Rev. B {\bf 68}, 184417, (2003).
\bibitem{SFT} {\it Electron Correlations and Magnetism in Narrow Band System}, edited by T.~Moriya (Springer, N.Y., 1981).
\bibitem{Kubler-book} J. K\"ubler, {\it Theory of Itinerant Electron Magnetism}, Clarendon Press (2000).
\bibitem{Vare} G. Varelogiannis, Phys.Rev.Lett. {\bf 91}, 117201, (2003).
\bibitem{Irisawa} K. Irisawa et al., Phys. Rev. B {\bf 70}, 214405, (2004).
\bibitem{blg} B. L. Gyorffy et al. J. Phys. F {\bf 15} 1337,(1985).
\bibitem{jh} J. Hubbard, Phys. Rev.B {\bf 20}, 4584, (1979).
\bibitem{hh} H. Hasegawa, J. Phys. Soc. Japan {\bf 46} 1504, (1979).
\bibitem{JBS+BLG} J. B. Staunton and B. L. Gyorffy, Phys.Rev.Lett. {\bf 69}, 371, (1992).
\bibitem{RDLM-FePt} J. B. Staunton et al., Phys.Rev.Lett. {\bf 93}, 257204, (2004).
\bibitem{Feynman} R. P. Feynman, Phys. Rev. {\bf 97}, 660, (1955).
\bibitem{RDLM-big} J. B. Staunton et al., Phys. Rev. B {\bf 74}, 144411, (2006).
\bibitem{Manuel} M. dos Santos Dias et al., Phys. Rev. B {\bf 83}, 054435 (2011).
\bibitem{Sandratskii} L. M. Sandratskii et al. Phys. Rev. B {\bf 76}, 184406, (2007).
\bibitem{Polesya} S. Polesya et al., Phys. Rev. B {\bf 82}, 214409, (2010). 
\bibitem{Nature} I. D. Hughes et al., Nature, {\bf 446}, 650, (2007).
\bibitem{TM-oxides} I. D. Hughes et al., New J. Physics, {\bf 10}, 063010, (2008).
\bibitem{IC1} A. Barcza et al. Phys.Rev.Lett. 104, 247202, (2010).
\bibitem{IC2} Z. Gercsi et al., Phys. Rev. B 83, 174403, (2011).
\bibitem{IC-2ndpaper} A.Barcza et al., in preparation (2012).
\bibitem{LTP} V. I. Valkov et al., Low Temp. Phys. {\bf 36}, 1064, (2010).
\bibitem{KKRCPA} G. M. Stocks et al., Phys. Rev. Lett. {\bf 41}, 339, (1978).
\bibitem{scf-kkr-cpa} D. D. Johnson et al., Phys. Rev. Lett. {\bf 56}, 2088, (1986).
\bibitem{KKR-review} H. Ebert et al., Rep. Prog. Phys. {\bf 74},096501, (2011). 
\bibitem{frozen} J. Harris, Phys. Rev. B {\bf 31}, 1770, (1985).
\bibitem{morrison_2008a} K. Morrison et al., Phys. Rev. B {\bf 78} 134418 (2008).
\bibitem{JMMM} Yu. D. Zavorotnev et al., J. Mag. Magn. Mat. {\bf 323}, 2808, (2011).
\end{thebibliography}
\end{document}